\documentstyle[12pt]{article}

 \title{Local microwave background radiation$^1$}

\author{Domingos S.L. Soares \\ Departamento de F\'{\i}sica, ICEx, UFMG 
--- C.P. 702 \\ 30123-970,  Belo Horizonte --- Brazil}

\date{\today}

\begin{document}
\maketitle

\hfill {\it `Big-Bang cosmology, the uncertain chain that links speculation to } 

\hfill {\it speculation in order to prove speculation.' }

\hfill Let it Bang,  Chronicles of Modern Cosmology

\hfill  -- D.S.L. Soares, unpublished \\

\hfill {\it `A hard rain's gonna fall means something's gonna happen.' }

\hfill No Direction Home: the soundtrack

\hfill  -- B. Dylan, 2005 \\

%
%
\addtocounter{footnote}{1}
\footnotetext{Further elaboration of a suggestion originally given in Soares 
2006a. } 
%

\begin{abstract}
An inquiry on a possible local origin for the Microwave Background Radiation 
is made. Thermal MBR photons are contained in a system called {\it magnetic 
bottle} which is due to Earth magnetic field and solar wind particles,
mostly electrons. Observational tests are anticipated.
\end{abstract}

\bigskip\bigskip\bigskip

\input epsf

\section{Introduction}
Cosmology is still a heavy-speculated field in spite of the enormous efforts 
on presumable cosmology-sensitive observations. In such an environment, 
scientists are not expected to make incisive statements unless they are 
supported by definitely secure evidence, both on the theoretical and 
experimental or observational sides. The cosmological standard model
does not fulfill the requirements of scientific method, therefore, cannot
be considered as a likely model for the {\it actual} universe: it relies
heavily on a number of {\it unknowns}, namely, inflaton field, baryonic
dark matter, non baryonic dark matter, dark energy, etc (see figures for dark 
components in Soares 2002). The concept of a
``cosmic" microwave background radiation (MBR) is introduced in the model
in a {\it ad hoc} fashion. In spite of that, it is even taken as a proof
of the model. But MBR may not be cosmic at all in the first place. Therefore 
there is a real necessity of investigating other causes or sources for it. The 
present paper considers a {\it local} origin for the radiation.

It is worthwhile mentioning two authoritative opinions on 
the  significance of the microwave background radiation in cosmology. Fred Hoyle 
(2001) states that 

\begin{quote}
``There is no explanation at all of the microwave background in the Big Bang 
theory. All you can say for the theory is that it permits you to put it in if 
you want to put it in. So, you look and it is there, so you put it in directly. 
It isn't an explanation."
\end{quote}

And Jean-Claude Pecker (2001) reaffirms:

\begin{quote}
``Actually, the 3 degree radiation, to me, has not a cosmological value. It is 
observed in any cosmology: in any cosmology you can predict the 3 degree 
radiation. So it is a proof of no cosmology at all, if it can be predicted  
of all cosmology."
\end{quote}

In this sense, also, it may prove in the future that the word
``cosmic" in the 1978 Nobel prize citation for A. Penzias and R. Wilson was
tendentiously premature: {\it ``for their discovery of {\bf cosmic}
microwave background radiation".}

The plan of the present paper is as follows. In section 2, the magnetic bottle 
scenario for a local MBR is presented and features of a related physical model are 
summarized. Section 3 discusses observational tests of a local MBR. In section 
4, final remarks are presented. 

\section{The Microwave Background Radiation and the magnetic bottle scenario }
Halton Arp in one of his 
books (Arp 1998, p. 237; see also Arp et al., 1990, p. 810) cites an 
authentic Fred Hoyle's aphorism: 

\begin{quote}
``A man who falls asleep on the top of a mountain and who awakes in a fog does not 
think he is looking at the origin of the Universe. He thinks he is in a fog."
\end{quote}

Let us then consider a {\it local} approach to the Microwave Background 
Radiation (MBR). Being freed from the 
``conceptual prison" of the Hot Big Bang Cosmology one may speculate on an earthly origin 
for the MBR. Earth's magnetosphere can be seen as a {\it magnetic bottle} whose 
walls are made by solar wind particles trapped along the magnetic lines of the 
Earth field. A minute fraction of Sun's light reflected by the Earth 
surface is caught within such a bottle and is thermalized through Thomson 
scattering on the bottle walls. The first consequence is that one would expect 
that the thermalized radiation should exhibit a {\it dipole anisotropy}, 
given the nature of Earth's  magnetic field. And that is precisely what was 
observed by the COBE satellite from its 900-km altitude orbit.   

Although WMAP, the 
{\it Wilkinson Microwave Anisotropy Probe}, sits far away 
from Earth, at the Lagrangean L2 point of the Sun-Earth system (see WMAP 
electronic page at the URL \\
{\tt http://map.gsfc.nasa.gov/m\_mm/ob\_techorbit1.html}, which means ab\-out 1.5 
million km from Earth, that is not enough for it to be released from the magnetic 
influence from Earth. 

It is located precisely 
and deep inside the bullet-shaped magnetopause, which extends to 1000 
times the Earth radius or more -- approximately 10 million km (see 
{\tt http://www-spof.gsfc.nasa.gov/Education/wmpause.html} 
for details of the magnetopause).

Figures 1 and 2 display the same geometry, as far as 
the Sun-Earth system is concerned. It is clear from the figures that as 
the Earth revolves about the Sun the Lagrangean point L2 -- thus WMAP -- 
sits all the time inside Earth's magnetopause.

\vfill

\begin{figure}[]
\begin{center}
\epsfxsize=12cm\epsfbox{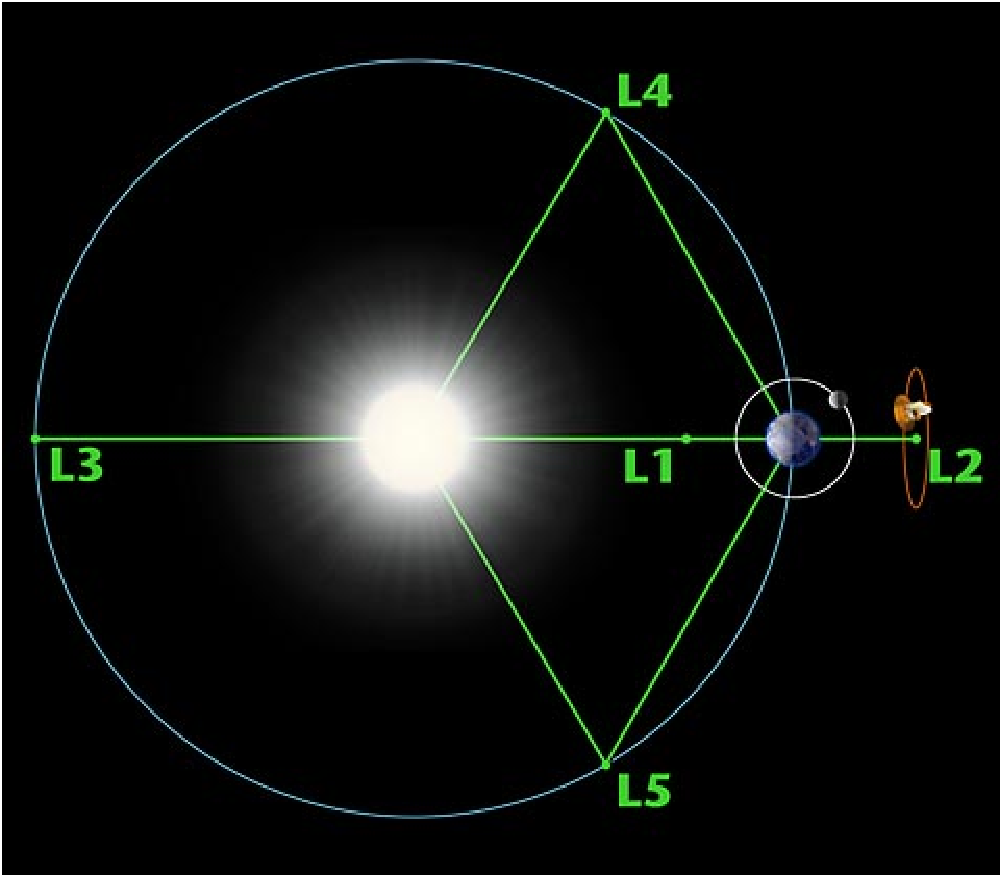}
\caption{Lagrangean points of the Sun-Earth system. 
WMAP satellite is shown at point L2. (Image credit: Wilkinson Microwave
Anisotropy Probe electronic page.) }
\end{center}
\end{figure}
%
%
\begin{figure}[]
\begin{center}
\epsfxsize=12cm\epsfbox{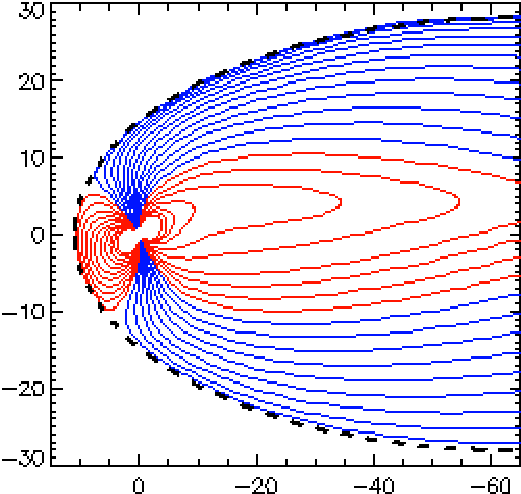}
\caption{ A view of Earth's magnetopause. The 
bullet-shaped magnetopause is always along the Sun-Earth direction (coordinates
in Earth radii). The magnetopause extends to up to 1000 Earth radii. L2 is 
inside the magnetopause at about 230 Earth radii. 
(Image credit: ``The Exploration of the Earth's Magnetosphere", an educational 
web site by David P. Stern and Mauricio Peredo.) }
\end{center}
\end{figure}

\vfill

The Earth magnetosphere has a complex structure with different electron densities 
in multiple layers around a neutral sheet at its mid-plane. The tail boundary
-- the magnetopause -- can reach 500-1000 Earth radii (Figure 2).

A simple model for the blackbody cavity may, in a first approximation, neglect 
anisotropies in the thermal spectrum (Soares 2006b). The magnetopause is 
modeled as a cylindrical cavity with its axis -- the z-direction -- running 
along the Sun-Earth direction, and $z=0$ at Earth's centre. Being $\phi$ the 
azimuth angle, the electron density $n(z, \phi)$, is a well-known observed 
quantity. Microwave photons are Thomson scattered inside the cavity till 
thermal equilibrium is attained in a time-scale much shorter than Earth's age. 
The precise source of the microwave photons is not critical since there 
are many possibilities. The most obvious is the long wavelength tail 
of the solar spectrum; the Earth itself 
might be another possibility (see below the discussion of radio thermal emission 
from solar system planets).

Hence, there is plenty of room for a coherent physical model of a blackbody cavity 
that generates the 3 K spectrum. Anisotropies are considered with a more realistic 
electron density distribution (Soares 2006b).

\section{ Observational tests  }
As long as one considers a local MBR, a plethora of observational tests come to 
light. Three major tests are discussed here.

\subsection{Non earthly MBR probe}
A straight consequence -- easily testable -- 
is that the background radiation from other ``magnetic bottles" -- other 
planets -- will be different, with a different thermal spectrum, possibly 
non thermal and even nonexistent. A probe orbiting another solar system planet 
like Mars, Venus, etc, would verify the hypothesis. 

\subsection{MBR anisotropy time variation}
The Earth magnetotail oscillates about its axis during the yearly revolution around 
the Sun by as much as 5 to 20 degrees (Eastman 2006). This introduces a measurable 
time variation on MBR anisotropies. An observational program that measures the 
MBR at different phases of Earth's orbit would detect such variations.

\subsection{ Planetary thermal glow}
The importance of radio thermal emission from planets is twofold. 
First, thermal emission may be the source of background radiation photons, and, 
second, the thermal glow may be the exterior manifestation of the background 
radiation itself. That is, the radiation which is interpreted as a ``cosmic" 
background radiation when measured from within the planetary environment -- the 
magnetic bottle -- is observed as a thermal glow from the outside.

There are many antecedents in observing thermal glows from planets in the 
radio-wave range. Mercury has a thermal 400 K glow and Venus was found 
to have an approximate 500 K glow by Mayer et al. (1958a). Radio emission from 
Mars and Jupiter at 3.15 cm and 9.4 cm are reported by Mayer et al. (1958b). 
A blackbody temperature of 210 K was found for Mars and 140 K for Jupiter. 

A reasonable prediction is that if one looks at the right wavelength range, one 
should be able to find the magnetic bottle signature of planetary
emission. Thus, the detection of Earth's 3 K thermal emission from 
the {\it outside} would be a strong indication of a local MBR.

\section{Concluding remarks}

\subsection{Martian Background Explorer (MABE)}
The next MBR anisotropy probe, NASA's Planck satellite, is scheduled for
launch in 2007. Again, it is planned to sit at Lagrangean L2 point, just
like WMAP (see briefing of Planck mission at\\
{\tt http://nssdc.gsfc.nasa.gov/database/MasterCatalog?sc=PLANCK}.

 It would be a great opportunity to test the validity of the
magnetic bottle scenario if Planck's observation site is moved to outside
the earthly environment. The immediate suggestion is a stationary point
on a Mars orbit, with the probe being obscured from solar radiation by the
planet, similar to the Sun-Earth-WMAP configuration.

Planck will measure, like WMAP did, background anisotropies. To measure the 
background radiation spectrum a COBE-like probe should be sent to Mars.

COBE -- the Cosmic Background Explorer -- was, in fact, "EARBE'', that is,
the Earthly Background Explorer. It measured the local background radiation. 
Whether "cosmic'', that is precisely the point. Thus, set up "MABE'' -- the 
Martian Background Explorer --, a replica of COBE except that placed at a 
low-altitude Martian orbit, equivalent to COBE's 900-km altitude orbit. 

Arm it with replicas of COBE's three instruments, FIRAS, DMR and 
DIRBE, and measure Martian background radi\-a\-tion spectrum and its 
a\-nis\-otro\-pies, just like COBE did on Earth.

Compare MABE's results with COBE's. 

The prediction is that the thermal background -- if it is indeed thermal --
will be totally different from the 3 K spectrum observed from Earth's
magnetic bottle.

\subsection{Historical note on the MBR}
Following the discovery of the MBR, Penzias \& Wilson published their findings 
in the 142nd volume of ApJ, in 1965. An accompanying paper, by Dicke et al. claimed
the {\it cosmic} nature of the phenomenon, establishing therefrom
the key foundation of the Big Bang cosmological model. They have in fact 
appropriated themselves of the discovery without leaving any room for other
tentative interpretations of the finding. Symptomatically --- as long as Dicke and 
collaborators were eager to take over the discovery in favor of their
ideas ---, their paper with {\it a possible} theoretical interpretation
was published {\it before} Penzias \& Wilson's report in ApJ. The observations
were referred to as {\it private communication}. Dicke et al. are at page
414 and Penzias \& Wilson at page 419 of ApJ's volume 142. The papers' 
titles also give the mood of both stories: Dicke et al. named theirs {\it Cosmic
Blackbody Radiation} --- a theory, of course --- while Penzias \& Wilson's paper 
was entitled {\it A Measurement of Excess Antenna Temperature at 4080 Mc/s} --- 
observations seeking for interpretation.  

They --- Penzias \& Wilson --- have discovered that, besides the smooth and
isotropic \emph{blue} background everyone could just {\it see}, the sky had 
also a smooth microwave background. Indeed, it was brilliantly confirmed
many years later by COBE, NASA's background explorer satellite.

But why, at that time, \emph{immediately} cosmic?

When in 1978 Penzias \& Wilson were granted the Nobel prize for their
discovery, the word ``cosmic'' was there, in the Nobel statement.
A political victory for the Big Bang cosmology. No science implied but
political strength acting on the Nobel committee. The first victory
in the political scenario of modern science achieved by the Big Bang
theory. At that time it was rather premature the ``cosmic'' attribute to the new 
finding. Nevertheless, it was sort of an ``official'' --- the Nobel committee --- 
approval of Big-Bang cosmologists' interpretation of the background radiation, 
and one which would become the theory's cornerstone. 

What seemed to be in scene was a {\it tour de force} between the Princeton 
group (Dicke et al.) and the Bell Labs scientists to get the 
credit for the great --- presumably 
cosmological --- discovery. A fight between giants: Princeton versus Bell. The 
Princeton group was in the end partially vindicated because the Nobel prize 
went to Penzias \& Wilson for the discovery but they managed to get the word 
``cosmic'' included in the Nobel statement: ``for their discovery of 
{\it cosmic} microwave background radiation.'' Without any doubt, a 
major triumph for a theory on the fighting against science-giant Fred Hoyle and 
collaborators with their steady-state cosmology.

More was to come though. 

In 2006, the descendants of the defeated group in the discovery of the MBR 
finally achieved their desired goal: the establishment's 
consecration of the Big-Bang theory and its dogmas. The Nobel prize in Physics of 
the year went to \emph{their} satellite: COBE.

The endeavor of COBE (Cosmic Background Explorer) was extraordinary
and the investigators which are responsible for it --- physicists
John Mather and George Smoot --- were quite important in the tremendous
scientific effort that involved hundreds of
technicians, engineers, physicists, astronomers, etc.

But it represents a technological development about something already
known, the Microwave Background Radiation (MBR), whose discovery has
already earned a Nobel prize in 1978.

It is clearly a development of technological nature and of experimental
improvement. There is nothing new as far as physics is concerned.
Would not anyone in sane conscience expect to find inhomogeneities in
the microwave background? Would it be a Nobel-like discovery to find them in the
isotropic blue background of our daylight sky?

There is no reason to believe that the MBR is of cosmological 
origin except if one is willing {\it to accept} a coordinated set of theoretical 
speculations with no firm observational bases whatsoever. 

The full story of the MBR imbroglio is still to be told. Let us wait because the 
best of it is certainly being nurtured. Crucial observational tests include:
\begin{enumerate}
\item time variability of the MBR on the scale of fraction of a solar year, and 
\item measurement of the MBR in another planetary environment.
\end{enumerate}
Both tests are unthinkable in the framework of a MBR with a \emph{cosmic} origin 
but are quite natural experiments from the point of view of a local origin 
for the MBR.

\section{References}
\begin{description}
\item Arp, H.C., Burbidge, G., Hoyle, F., Narlikar, J.V., Wickramasinghe, N.C. 
1990, Nature, 346, 807
\item Arp, H. 1998, {\it Seeing Red: Redshifts, Cosmology and Academy,} Apeiron, 
Montreal
\item Eastman, T. 2006, private communication; see also \\
{\tt http://www.plasmas.org/space-plasmas.htm}
\item Hoyle, F. 2001, in {\it Universe, The Cosmology Quest}, DVD directed 
by Randall Meyers, A Floating World Films production
\item Mayer, C.H., McCullough, T.P., Sloanaker, R.M. 1958a, ApJ, 127, 1
\item Mayer, C.H., McCullough, T.P., Sloanaker, R.M. 1958b, ApJ, 127, 11
\item Pecker, J.-C. 2001, in {\it Universe, The Cosmology Quest}, DVD directed 
by Randall Meyers, A Floating World Films production
\item Soares, D.S.L. 2002, {\it Do we live in an anthropic universe?},\\  
arXiv:physics/0209094 
\item Soares, D.S.L. 2006a, {\it Sandage versus Hubble on the reality of the 
expanding universe}, arXiv:physics/0605098
\item Soares, D.S.L. 2006b, in preparation

\end{description}

\end{document}